\documentclass[prl,twocolumn,showpacs]{revtex4}

\usepackage{graphicx}
\usepackage{amsmath}

\usepackage{amsmath}
\usepackage{amsfonts}

\usepackage{pstricks,pst-fill,pst-text,pst-node,pstricks-add}
\usepackage{psfrag}

\newcommand{\Ket}[1]{\left|#1  \right>}

\newcommand{\Braket}[1]{\left<#1  \right>}

\begin{document}

\title{The puzzle of bulk conformal field theories at central charge $c=0$}

\author{Romain Vasseur$^{1,2,5}$, Azat Gainutdinov$^{1,5}$, Jesper Lykke Jacobsen$^{2,3,5}$ and Hubert Saleur$^{1,4,5}$}
\affiliation{${}^1$Institut de Physique Th\'eorique, CEA Saclay,
91191 Gif Sur Yvette, France}
\affiliation{${}^2$LPTENS, 24 rue Lhomond, 75231 Paris, France}
\affiliation{${}^3$Universit\'e Pierre et Marie Curie, 4 place Jussieu, 75252 Paris, France}
\affiliation{${}^4$Department of Physics,
University of Southern California, Los Angeles, CA 90089-0484}
\affiliation{${}^5$Institut Henri Poincar\'e, 11 rue Pierre et Marie Curie, 75231 Paris Cedex 05, France}

\date{\today}

\begin{abstract}

  Non-trivial critical models in 2D with central charge $c=0$ are
  described by Logarithmic Conformal Field Theories (LCFTs), and
  exhibit in particular mixing of the stress-energy tensor with a
  ``logarithmic'' partner under a conformal transformation. This
  mixing is quantified by a parameter (usually denoted $b$),
  introduced in [V.~Gurarie, Nucl.\ Phys.\ B {\bf 546}, 765 (1999)].
  The value of $b$ has been determined over the last few years for the
  boundary versions of these models: $b_{\rm perco}=-\frac{5}{8}$ for
  percolation and $b_{\rm poly} = \frac{5}{6}$ for dilute
  polymers. Meanwhile, the existence and value of $b$ for the bulk
  theory has remained an open problem. Using lattice regularization
  techniques we provide here an ``experimental study'' of this
  question. We show that, while the chiral stress tensor has indeed a
  single logarithmic partner in the chiral sector of the theory, the
  value of $b$ is not the expected one: instead, $b=-5$ for {\sl both}
  theories.  We suggest a theoretical explanation of this result using
  operator product expansions and Coulomb gas arguments, and discuss
  the physical consequences on correlation functions.  Our results
  imply that the relation between bulk LCFTs of physical interest and
  their boundary counterparts is considerably more involved than in
  the non-logarithmic case.

\end{abstract}


\maketitle

\paragraph{Introduction.}

Two-dimensional (2D) logarithmic conformal field theories (LCFTs) with
central charge $c=0$ describe the scaling limit of geometrical
problems such as polymers (self-avoiding walks) or percolation (see
{\it e.g.}~\cite{SaleurPoly}).  They also play a fundamental role in
the study of phase transitions in systems with quenched disorder
\cite{Cardylog, GurarieLudwig,GurarieLudwig2}: for instance, the long
sought-after conformal field theory describing the plateaux transition
in the integer quantum Hall effect should be of this type.
Other physical applications range from non-equilibrium
systems~\cite{Raisepeels} to aspects of the AdS/CFT correspondence
(see {\it e.g.}~\cite{Smodel}) and description of super-symmetric sigma models beyond the topological sector~\cite{FLN}.

Progress in this field since the pioneering papers
\cite{RozanskySaleur,Gurarie} has been difficult. Chiral (i.e., boundary) LCFTs are
slowly getting under control thanks to recent progress on the abstract
study of non semi-simple modules of the Virasoro algebra---the
so-called staggered modules~\cite{KytolaRidout}. Consistent results
have also been obtained from lattice approaches~\cite{RS3,PRZ}. It is
now well-accepted that chiral LCFTs are characterized by a complicated
structure of such staggered modules, and an infinity of numbers
called indecomposability parameters, or logarithmic couplings.  One of
these numbers, called $b$~\cite{Gurarie99,GurarieLudwig2}, is of
particular interest as it encodes the structure of the module of the
stress-energy tensor in $c=0$ theories. Numerical~\cite{DJS,VJS} and
theoretical~\cite{MathieuRidout} arguments lead to the values
$b_{\rm perco} = -\frac{5}{8}$ for percolation [$Q \to 1$ state
Potts model] and $b_{\rm poly} = \frac{5}{6}$ for dilute polymers
[$n \to 0$ dilute O($n$) model].

While chiral LCFTs are thus fairly well understood, much remains to be
done to understand the structure of bulk theories.  Attempts to
construct non-chiral theories at $c=0$ (see {\em
  e.g.}~\cite{Flohr,Triplet}) often
exhibit unwanted features such as degenerate or non $SL(2,\mathbb{C})$
invariant ground states which should not occur in, for instance,
percolation.

Part of the progress in the chiral case originates from considering
lattice models with open boundary conditions, and
observing that the indecomposable features of the chiral (Virasoro)
algebra appearing in the scaling limit are similar (for more
accurate, mathematical statements, see~\cite{RS2}) to those occurring,
in finite size, on the lattice. This suggests that eventually, LCFTs
may be solved by a careful exercise in the representation theory of
the (associative) algebras satisfied by the local energy terms (such
as the Temperley-Lieb algebra~\cite{PPMartin}). The non-chiral or bulk case
corresponds, on the lattice, to periodic boundary conditions, which is
rather difficult mathematically~\cite{GRS}. One can nevertheless
use lattice algebraic techniques to investigate aspects of the
simplest modules under the full left and right Virasoro algebras
present in this case.  The full results will be discussed elsewhere
\cite{VGJRS}, but, as a first step, we present in this Letter the
results of the corresponding numerical measure of the parameters $b$
for both percolation and polymers with periodic boundary
conditions (PBC). The outcome is quite unexpected: both theories turn
out to have the same $b$, which moreover differs from that of the open
(chiral) case.

\paragraph{b parameter.}
We begin by defining the $b$ number, first introduced
by Gurarie ~\cite{Gurarie99,GurarieLudwig, GurarieLudwig2}.  We consider
the (non-chiral) LCFTs that describe the scaling limit of polymers or
percolation. Within these theories, the stress-energy tensor $T(z)=
L_{-2} I$ is mixed into a Jordan cell with its logarithmic partner
$t(z,\bar{z})$ such that in the basis $(T,t)$, the generator of the
scale transformation reads
\begin{equation}\label{eq_Jcell}
\displaystyle L_0+{\bar L_0} =  \left( \begin{array}{cc} 2 & 2  \\ 0 & 2  \end{array} \right).
\end{equation}
We will show later (see also~\cite{KoganNichols}) that the
non-diagonalizable term can be decomposed as $L_0 t = 2t + T$ and
$\bar{L_0} t = T$, so that, unlike $T(z)$ which is purely holomorphic,
the field $t(z,\bar{z})$ has a non-trivial antiholomorphic part
$\bar{\partial} t \neq 0$.  Using this normalization, we define the
number $b$ as
\begin{equation}
\displaystyle b = \Braket{T | t},
\end{equation}
where we have used the bilinear Virasoro form. Note also that $L_{1} t=0$ and
$L_{2} t = b I$. Using these relations, global conformal invariance
fixes \cite{GurarieLudwig} the form of the two-point functions to be
\begin{subequations}\label{eqCorr}
\begin{eqnarray}
\left\langle T(z) T(0)\right\rangle &=& 0 \\
\left\langle T(z) t(0,0)\right\rangle &=& \frac{b}{z^{4}} \\
\left\langle t(z,\bar{z}) t(0,0)\right\rangle &=& \frac{\theta - 2 b \log \left| z \right|^2} {z^{4}} \,,
\end{eqnarray}
\end{subequations}
where $\theta$ is an irrelevant constant that can absorbed into a
redefinition of the fields, whereas $b$ is a fundamental number
characterizing the logarithmic pair $(T,t)$.  We have similar
equations for the antiholomorphic counterpart
($\bar{T},\bar{t}\ldots$), with also $\langle t \bar{t}\rangle =
\langle T \bar{t}\rangle = 0 $.

\paragraph{Lattice loop models.}

\begin{figure} 
\includegraphics[width=8.0cm]{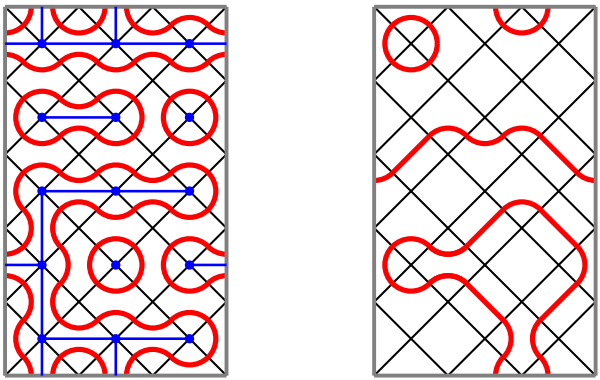}
\caption{Example of dense (left panel) and dilute (right panel) loop
  configurations. We use periodic boundary conditions in both
  directions so that the loops actually live on a torus. We also show
  the corresponding FK clusters for the dense loop model. Each closed loop
  has a weight $n$.}
  \label{FigLoop}
\end{figure}

Our aim is to measure $b$ in percolation and dilute
polymers with PBC. An essential difficulty consists
in defining the models; since, for instance, percolation presents many
non-local (or apparently non-local) aspects, the consideration of
different geometrical features might lead to different, mutually
inconsistent, field theoretic descriptions. A safe way out is to use
supersymmetry~\cite{ReadSaleur}, but to keep things simple, we
consider---as is quite standard---the 
observables encoded in a gas of loops, where each loop carries
a weight $n$.

It is well-known how to rewrite the $Q$-state Potts model in
terms of Fortuin-Kasteleyn (FK) clusters (see 
{\it e.g.}~\cite{Baxter}), whose hulls are {\em dense} loops with $n =
\sqrt{Q}$. Percolation then arises as the $n \to 1$ limit.  We choose
PBC in both space and imaginary time directions. The corresponding
correlations can be obtained using a transfer matrix, or, in the very
anisotropic limit, a quantum Hamiltonian in 1+1D, $H = -
\sum_{i=1}^{2N} e_i$, where $L=2N$ is the length of the system in the space direction
and the $e_i$ are Temperley-Lieb (TL) generators. These are represented by the contraction $e_i =
\psset{xunit=3mm,yunit=3mm} \begin{pspicture}(0,0)(1,1)
  \psellipticarc[linecolor=black,linewidth=1.0pt]{-}(0.5,1)(0.355,0.355){180}{360}
  \psellipticarc[linecolor=black,linewidth=1.0pt]{-}(0.5,0)(0.355,0.355){0}{180}
\end{pspicture}
$
 of the sites $i$ and $i+1$ on the cylinder lattice in Fig.~\ref{FigLoop}. 
The
different sectors of the ``Hilbert'' space of the Hamiltonian are
labeled by the number of through-lines $2j$ propagating along the
imaginary time direction.
 Examples of
states in the $j=1$ sector on $L=6$ sites are given by
$\Ket{\alpha} = \Ket{
\psset{xunit=2mm,yunit=2mm}
\begin{pspicture}(0,0)(5.5,1)
 \psellipticarc[linecolor=black,linewidth=1.0pt]{-}(1.5,1.0)(1.5,1.42){180}{360}
 \psellipticarc[linecolor=black,linewidth=1.0pt]{-}(1.5,1.0)(0.5,0.71){180}{360}
 \psline[linecolor=black,linewidth=1.0pt]{-}(4,-0.50)(4,1)
 \psline[linecolor=black,linewidth=1.0pt]{-}(5,-0.50)(5,1)
\end{pspicture}
}$
or $\Ket{\beta} = \Ket{
\psset{xunit=2mm,yunit=2mm}
\begin{pspicture}(0,0)(5.5,1)
 \psellipticarc[linecolor=black,linewidth=1.0pt]{-}(0,1.0)(1.5,1.42){270}{360}
 \psellipticarc[linecolor=black,linewidth=1.0pt]{-}(0,1.0)(0.5,0.71){270}{360}
 \psline[linecolor=black,linewidth=1.0pt]{-}(2,-0.50)(2,1)
 \psline[linecolor=black,linewidth=1.0pt]{-}(3,-0.50)(3,1)
 \psellipticarc[linecolor=black,linewidth=1.0pt]{-}(5,1.0)(1.5,1.42){180}{270}
 \psellipticarc[linecolor=black,linewidth=1.0pt]{-}(5,1.0)(0.5,0.71){180}{270}
\end{pspicture}
}$ imposing PBC.
 The action of the $e_i$ can produce winding of through-lines around the cylinder which we replace
 by a weight $1$; note also that non-contractible
and contractible loops have the same weight $n=1$~\cite{Jones,GL}.
Under the action of $H$, the number
of through-lines can only decrease, since
$e_i$
%
corresponds to a contraction.
$H$ is thus a lower block-triangular matrix which corresponds in the scaling limit to $L_0+\bar{L}_{0}$ and we
expect to see Jordan cells on a finite lattice.
In order to study Jordan cells for percolation, we use the
same prescription as in the open case (see~\cite{DJS}).

To describe dilute polymers, we use the integrable version of the
$O(n\rightarrow0)$ model on the square lattice~\cite{Onweights} which
yields a model of {\em dilute} loops with fugacity $n=0$. We can write the
Hamiltonian as a sum of local densities as before, $H=-\sum_i e_i$,
where the operator $e_i$ is now more complicated (see {\it e.g.}~\cite{VJS}).
In this case $j$ denotes the number of through-lines.
The states are the same as in the dense case, except
that we now allow for empty sites. Examples of configurations are shown
in Fig.~\ref{FigLoop}.

\paragraph{Lattice Jordan cells and scalar products.}

Both models have trivial partition functions, so their scaling limit
is described by conformal theories with $c=0$.  The
identity operator corresponds to the groundstate of $H$, henceforth
denoted $\Ket{0^{(N)}}$, and can be found in the vacuum ($j=0$) sector. The
state corresponding to the stress-energy tensor $T(z)$ is the only
state with conformal weights $(h,\bar{h})=(2,0)$ in the vacuum sector.
The scaling dimension $\Delta = h+\bar{h}$ of a given
excitation $\phi$ with energy $E_{\phi}(N)$ can be measured
numerically from the scaled energy gap, $E_{\phi}(N)-E_{0}(N)
= \frac{2\pi v_F}{2N} \Delta + \dots$, where $E_{0}(N)$ is the energy
of the groundstate and $v_F$ is the Fermi velocity.  Bethe ansatz
results~\cite{PottsBethe,OnBethe} yield $v_F=\frac{3 \sqrt{3}}{2}$ for
percolation and $v_{F} = \frac{8}{3}$ for dilute polymers.
Both the Hamiltonian and the
transfer matrix commute with the operator $u^{2}$ that translates the
sites two units to the right.
The restriction to a given conformal spin $s=h-\bar{h}$ can be done
using this translational invariance.
We can build eigenspaces for $u^2$
and introduce a momentum $k$ to characterize a given
sector. 
It can
be argued that the states with conformal spin $s$ belong to
the sector with momentum $k=\frac{2 \pi}{N} s$. Using these results,
one can readily extract the states with conformal weights
$(h,\bar{h})=(2,0)$ from the spectrum.  We find numerically that the
state $\Ket{T^{(N)}}$, which goes to $\Ket{T}=T(z\to0)\Ket{0}$ in the scaling limit $N
\to \infty$ [up to normalization issues to be discussed below], is
mixed into a rank-$2$ Jordan cell with its ``logarithmic partner''
$\Ket{t^{(N)}}$. Note that $\Ket{t^{(N)}}$ belongs to the
$j=2$ sector.  We normalize the states so that the Hamiltonian in the
basis $\bigl(\Ket{T^{(N)}},\Ket{t^{(N)}}\bigr)$ reads
\begin{equation}
\displaystyle H - E_{0}(N) {\bf 1} =  \frac{2 \pi v_F}{2N} \left( \begin{array}{cc} \Delta^{(N)} & 2  \\ 0 & \Delta^{(N)}  \end{array} \right),
\end{equation}
where $\Delta^{(N)}= \frac{2N}{2\pi v_F} (E_{T}(N)-E_{0}(N))$. In
order to measure $b=\Braket{T|t}$, we also need a ``scalar product''
which goes to the Virasoro bilinear form in the scaling limit. This
issue was already discussed in Refs.~\cite{DJS,VJS} and is not
modified by the PBC: the natural definition of the scalar product for
such loop models corresponds to gluing the mirror image of the
first state on top of the second one.

\paragraph{Virasoro algebra regularization.}

It could be tempting to define a lattice version of $b$ as the scalar
product $\Braket{t^{(N)}| T^{(N)}}$.  However, the Jordan cell in the
Hamiltonian is invariant under a simultaneous rescaling of the states
$\Ket{T^{(N)}}$ and $\Ket{t^{(N)}}$. Therefore, we need to carefully
normalize $\Ket{T^{(N)}}$ so that it goes {\em precisely} to the
stress-energy tensor in the scaling limit \cite{DJS}; since
$\Braket{T|T}=0$, the problem is non-trivial, and has in fact hindered
for many years the numerical determination of $b$.  Like in the chiral
case~\cite{VJS}, we control normalization by using a lattice
regularization $L^{(N)}_n$, $\bar{L}^{(N)}_{n}$ of the Virasoro
generators~\cite{KooSaleur}. Introducing the notation $H^{(N)}_n =
L^{(N)}_n + \bar{L}^{(N)}_{-n}$, arguments such as those in~\cite{KooSaleur}
lead, for both our dense and dilute loop models, to
%
\begin{equation}\label{eqKooSaleur}
\displaystyle H^{(N)}_n = -\frac{N}{\pi v_F} \sum_{j=1}^{2N} \mathrm{e}^{inj \pi /N} \left( e_i - e_{\infty} \right) + \frac{c}{12} \delta_{n,0},
\end{equation}
%
where $e_\infty = 1$ ({\it resp.} $e_\infty = \sqrt{2}$) for
percolation ({\it resp.} dilute polymers) is the groundstate energy density.

\paragraph{Numerical results.}

\begin{table}
\begin{center}
\begin{tabular}{|c|c|c|c|}
\cline{1-4}
  \multicolumn{2}{|c|}{Percolation} & \multicolumn{2}{|c|}{Dilute Polymers} \\ \cline{1-4}
  $L = 2 N$ & $b^{(N)}$ & $L = 2 N$ & $b^{(N)}$ \\
  \hline
  10 & -4.33296  &   &  \\
  12 & -4.55078  &   & \\
  14 & -4.68234  & 10  & -4.17430 \\
  16 & -4.76634  & 12  & -4.38064 \\
  18 & -4.82256  & 14  & -4.52458 \\ 
  20 & -4.86168  &  & \\ 
  22 & -4.88978  &   & \\
  \hline
  $\infty$ & -5.00 $\pm$ 0.01  & $\infty$ & $\simeq$ -5 \\
  \hline
  \hline
  Exact & $-5$ & Exact & $-5$\\
  \hline
\end{tabular}
\end{center}
\caption{Measure of the indecomposability parameter $b$ for percolation
 and dilute polymers.}
  \label{Bulkb}
\end{table}

Gathering all these (admittedly elaborate) pieces, we finally give a
lattice expression that goes to $b$ in the scaling limit~\cite{VJS}
\begin{equation}\label{eq_bnum}
\displaystyle b^{(N)} = \frac{ \left| \Braket{t^{(N)}|H^{(N)}_{-2} |0^{(N)}} \right|^2 }{ \Braket{t^{(N)}|T^{(N)}} },
\end{equation}
where we have normalized the groundstate so that
$\Braket{0^{(N)}|0^{(N)}}=1$. 
 Numerical
results for percolation
%
%
and dilute polymers are gathered in
Tab.~\ref{Bulkb}.  Both theories seem to have $b=-5$, although the
precision is obviously much better for percolation as the Hilbert
space is much smaller. Note that we also measured $b^{(N)}$ in the
$\mathfrak{sl}(2|1)$-supersymmetric representation of
percolation~\cite{SQHE,ReadSaleur}, with the same result~\cite{VGJRS}. 
Our value of $b$ thus also applies to the CFT describing the Spin Quantum Hall Transition~\cite{SQHE}. 

The value $b=-5$ is quite surprising, since the standard
argument identifies to a large extent the chiral sector of the bulk
theory with the chiral theory~\cite{GurarieLudwig}, and thus one would
expect to recover the standard values $b = -\frac{5}{8}$ or
$\frac{5}{6}$.
We now turn to an analytical argument that allows us to
derive the value $b_{\rm bulk}=-5$ exactly.

\paragraph{OPE approach.}
It was shown~\cite{VJS} that by pushing further the ideas in~\cite{KoganNichols,GurarieLudwig2,Cardylog}
one can predict the value of indecomposability parameters in the chiral
case using operator product expansions (OPEs). The idea is that the identity channel of
the OPE of a primary operator $\Phi_{h,\bar{h}}$ (with conformal
weights $(h,\bar{h})$) with itself is ill-defined as $c \rightarrow 0$
\begin{equation}\label{eq_OPE1}
\displaystyle \Phi_{h,\bar{h}}(z,\bar{z}) \Phi_{h,\bar{h}}(0,0) \sim \frac{a_\Phi}{z^{2h}\bar{z}^{2\bar{h}}}  \left[1 + \frac{2 h}{c} z^2 T(0)+ \dots \right],
\end{equation}
where $a_\Phi$ is a coefficient and we focused on the holomorphic sector on the right-hand side (a
similar reasoning applies to $\bar{T}$). The divergence as
$c\rightarrow 0$ can be cured if there exists a primary
field $X(z,\bar{z})$ with conformal weights $(h_X(c), \bar{h}_X(c))
\rightarrow (2, 0)$ that ``collides'' with the stress tensor at $c=0$.
One can show that a combination of $X(z,\bar{z})$ and $T(z)$ becomes
the logarithmic partner $t(z)$ of $T(z)$ at $c=0$, with correlation 
functions like eqs.~\eqref{eqCorr} consistent with a value of $b$ given by
$b=\lim_{c\rightarrow0} - \frac{c/2}{h_X(c)-2}=\lim_{c\rightarrow0} -
c/2\bar{h}_X(c)$. We assumed that those limits exist and are equal; 
this turns out to be satisfied in the following.

\paragraph{Coulomb Gas.}
In the remainder of this Letter, we shall denote the bulk
fields $\Phi_{r,s}(z) \otimes \Phi_{r',s'}(\bar{z})$ using Kac labels,
meaning that the conformal dimensions fit into the Kac table as
$(h=h_{r,s},\bar{h}=h_{r',s'})$, where we use the parametrization
$h_{r,s}=\frac{\left[r (x+1)-s x \right]^2-1}{4x(x+1)}$ and
$c=1-\frac{6}{x(x+1)}$ (our models with $c=0$ thus correspond to
$x=2$).  The crucial step is now to identify the field $X(z,\bar{z})$,
since $h_X(c)$ then determines $b$.  In
the standard approach to this problem
~\cite{GurarieLudwig,GurarieLudwig2}, one looks for $X$ among
degenerate fields such as $\Phi_{3,1}$ or $\Phi_{1,5}$ in the chiral
case.  Considerations on the underlying lattice model however gives
an unambiguous answer ~\cite{FSZ}.  First, recall that the dense loop
model renormalizes towards a Coulomb Gas (CG)~\cite{NienhuisLoop} with
coupling constant $g \in [2,4]$ given by
$Q = 2(1+ \cos \frac{\pi g}{2})$---recall that $n=\sqrt{Q}$ is the loop weight.
The associated
central charge reads $c^{(Q)} = 1 - \frac{3(g-4)^2}{2g}$, and
percolation corresponds to $Q=1$, $g=\frac{8}{3}$.  Similarly
the $O(n)$ model can be expressed as dilute loops that renormalize
towards a Coulomb gas with coupling constant $g \in [1,2]$ given by
$ n = -2 \cos \pi g$, so dilute polymers have $g=\frac{3}{2}$.  The
central charge in this case reads $c^{(n)} = 1 - \frac{6(g-1)^2}{g}$.
The operator content of the loop models can then be worked out from
their partition function on a torus~\cite{FSZ,ReadSaleur}. 
The candidate for the 
field $X(z,\bar{z})$  must be a spin-2 operator, and knowledge from
the boundary case, from our numerics, and from replica considerations
in the case of the $O(n)$ model~\cite{Cardylog} suggests that $X(z,\bar{z})$
has to be a 2-leg ({\it resp}. 4-leg) operator for dilute polymers ({\it resp}. percolation).
We find that it has conformal weights ($h=\frac{(2 + g)^2}{4g} + \frac{c-1}{24}$ ,
$\bar{h}=\frac{(2 - g)^2}{4g} + \frac{c-1}{24}$) in both cases. 
Moreover this is the {\it only  possible} choice in the CG that
would yield exponents $(2,0)$ at $c=0$. In
terms of Kac labels, we can formally identify this field as
$X(z,\bar{z}) = \Phi_{1,-2}(z) \otimes \Phi_{1,2}(\bar{z})$, which is non-degenerate. We finally
compute $b$ using the continuation of the critical
exponents of $X$, finding
\begin{subequations}
\begin{eqnarray}
\displaystyle b_{\mathrm{perco}} &=&-\lim_{g \rightarrow \frac{8}{3}} \dfrac{c^{(Q)} /2}{\frac{(2 - g)^2}{4g} + \frac{c^{(Q)}-1}{24}} = -5, \\
\displaystyle b_{\mathrm{poly}} &=& -\lim_{g \rightarrow \frac{3}{2}} \dfrac{c^{(n)} /2}{\frac{(2 - g)^2}{4g} + \frac{c^{(n)}-1}{24}} = -5.
\end{eqnarray}
\end{subequations}
in perfect agreement with our numerical results.

\paragraph{Physical discussion.} We believe that
our result is quite crucial for the general structure of bulk LCFTs at
$c=0$. In particular, the value $b=-5$ should have profound
consequences on the bulk correlation
functions~\cite{GurarieLudwig2,SC}. Indeed, the solution of the
differential equations for the four-point correlation function of the
degenerate bulk field $\Phi_{2,1}(z,\bar{z}) \equiv \Phi_{2,1}(z) \otimes
\Phi_{2,1}(\bar{z})$ (with conformal weights
$(\frac{5}{8},\frac{5}{8})$) or 
$\Phi_{1,3}(z,\bar{z}) \equiv \Phi_{1,3}(z) \otimes
\Phi_{1,3}(\bar{z}) $ (with conformal weights
$(\frac{1}{3},\frac{1}{3})$), and the resulting conformal blocks lead
unambiguously to the values $b_{\rm poly}=\frac{5}{6}$ and $b_{\rm perco}=-\frac{5}{8}$
respectively (this is one of the arguments used
in~\cite{GurarieLudwig2} to predict the value of $b$ for percolation
and polymers), {\em provided that} one ignores normalization issues in the bulk. That
$b$ takes none of these values forces one to conclude that, within the
$c=0$ conformal field theory describing percolation ({\it resp.}
polymers), the four point correlations of the energy operator $\Phi_{2,1}$
({\it resp.} $\Phi_{1,3}$)~\cite{DF} must be set equal to zero exactly. This is
compatible with the known---but maybe underestimated---fact that one
cannot normalize these correlation functions to render them
non-trivial and without divergences~\cite{Cardylog}. Note also that we were able to
show from both algebraic and numerical arguments that the energy
operator $\Phi_{2,1}$ ({\it resp.} $\Phi_{1,3}$) is mixed into a
rank-2 Jordan cell, and hence has a zero norm-squared. One can deduce that
{\em all} its correlation functions 
are actually vanishing~\cite{Flohr2}, and thus
irrelevant for the determination of $b$. Note that
the vanishing of the multi-point correlation functions of the energy
does not preclude other correlators involving the energy from taking
non-zero values. 
The computation of non-vanishing correlation functions in general
nevertheless appears unreachable.  Indeed, since $b=-5$
and occurs, from our OPE approach, because of mixing
with operators having {\em negative} Kac labels,
progress would presumably require us to control the four-point
functions of such operators---a situation for which the differential
equation or CG approaches are known to fail.

Note that the crucial point of our calculation is the proper identification
of the field $X(z,\bar{z})$ among primary operators given by the CG.
We emphasize that within this natural ``minimal'' 
geometrical formulation of the problem, there is no bulk counterpart
of the boundary fields $\Phi_{3,1}$ or $\Phi_{1,5}$ that played
the role of $X$ in the chiral case. Instead, we have shown that $X(z,\bar{z})$ must
be a non-Kac operator with negative Kac labels, thus yielding 
a value of $b$ different from what was expected from, {\it e.g.}, Refs.~\cite{GurarieLudwig2,SC}.
The possibility of having a value of $b$ in the bulk different from the boundary
case was suggested in Ref.~\cite{SC}. We emphasize that our geometrical bulk theory admits a
{\it unique} value of $b$~\footnote{In particular, there is no contradiction with the argument given in Appendix A of Ref.\cite{GurarieLudwig}.}.
Note however that a ``non-minimal'' formulation---e.g., involving further geometrical observables \cite{Zinn,DW}---might eventually lead
to a more complicated Jordan structure for $T(z)$.
 Finally, we remark that $\frac{2}{b_{\rm bulk}}=\frac{1}{b_{\rm perco}} +
\frac{1}{b_{\rm poly}}$; this is an
intriguing observation indeed, since $b^{-1}$ is the parameter that appears multiplicatively in
correlation functions and OPEs.

In conclusion, we see that the complete picture of the bulk theory of
percolation (or polymers) is more complicated than expected, cannot be
seen as a simple gluing of two chiral theories, and seems in fact to
have very little to do with the nice understanding gained over the
last few years in the boundary (chiral) case. Further understanding follows
from a detailed algebraic analysis, which will
appear elsewhere~\cite{VGJRS}.

\smallskip

\paragraph{Acknowledgments.}
We gratefully acknowledge stimulating discussions and collaboration on
related matters with N.~Read. We also thank J.~Cardy for interesting discussions.
This work was supported by the Agence
Nationale de la Recherche (grant ANR-10-BLAN-0414: DIME). The work of AG was supported in 
part by Marie-Curie IIF fellowship, the RFBR grant 10-01-00408 and the RFBRCNRS grant 09-01-93105.

\end{document}